\documentclass[aip,jmp,preprint,onecolumn,amsmath,amssymb,groupedaddress]{revtex4-1}

\usepackage{graphicx}
\usepackage{dcolumn}
\usepackage{bm}
\usepackage{hyperref}
\usepackage[mathlines]{lineno}
\usepackage{braket}
\usepackage{lipsum}
\usepackage{alltt}
\usepackage{multirow}

\makeatletter
\allowdisplaybreaks[2]

\newcommand{\e}{{\rm e}}
\newcommand{\tr}{{\mathrm{tr}}}

\newcommand{\dd}{\,\mathrm{d}}

\newmuskip\pFqmuskip
\newcommand*\pFq[6][8]{%
  \begingroup
  \pFqmuskip=#1mu\relax
  \mathchardef\normalcomma=\mathcode`,
  \mathcode`\,=\string"8000
  \begingroup\lccode`\~=`\,
  \lowercase{\endgroup\let~}\pFqcomma
  {}_{#2}F_{#3}{\left[\genfrac..{0pt}{}{#4}{#5};#6\right]}%
  \endgroup}
\newcommand{\pFqcomma}{{\normalcomma}\mskip\pFqmuskip}

\begin{document}
\title{On the exact variance of Tsallis entropy in a random pure state}
\author{Lu Wei}
\email{luwe@umich.edu}
\affiliation{Department of Electrical and Computer Engineering \\ University of Michigan-Dearborn, MI 48128, USA}
\date{\today}

\begin{abstract}
Tsallis entropy is a useful one-parameter generalization of the standard von Neumann entropy in information theory. We study the variance of Tsallis entropy of bipartite quantum systems in a random pure state. The main result is an exact variance formula of Tsallis entropy that involves finite sums of some terminating hypergeometric functions. In the special cases of quadratic entropy and small subsystem dimensions, the main result is further simplified to explicit variance expressions. As a byproduct, we find an independent proof of the recently proved variance formula of von Neumann entropy based on the derived moment relation to the Tsallis entropy.
\end{abstract}

\maketitle

\section{Introduction}
We consider a composite quantum system consisting of two subsystems $A$ and $B$ of Hilbert space dimensions $m$ and $n$, respectively. The Hilbert space $\mathcal{H}_{A+B}$ of the composite system is given by the tensor product of the Hilbert spaces of the subsystems, $\mathcal{H}_{A+B}=\mathcal{H}_{A}\otimes\mathcal{H}_{B}$. The random pure state (as opposed to the mixed state) of the composite system is written as a linear combination of the random coefficients $x_{i,j}$ and the complete basis $\left\{\Ket{i^{A}}\right\}$ and $\left\{\Ket{j^{B}}\right\}$ of $\mathcal{H}_{A}$ and $\mathcal{H}_{B}$, $\Ket{\psi}=\sum_{i=1}^{m}\sum_{j=1}^{n}x_{i,j}\Ket{i^{A}}\otimes\Ket{j^{B}}$. The corresponding density matrix $\rho=\Ket{\psi}\Bra{\psi}$ has the natural constraint $\tr(\rho)=1$. This implies that the $m\times n$ random coefficient matrix $\mathbf{X}=(x_{i,j})$ satisfies
\begin{equation}\label{eq:pcv}
\tr\left(\mathbf{XX}^{\dag}\right)=1.
\end{equation}
Without loss of generality, it is assumed that $m\leq n$. The reduced density matrix $\rho_{A}$ of the smaller subsystem $A$ admits the Schmidt decomposition $\rho_{A}=\sum_{i=1}^{m}\lambda_{i}\Ket{\phi_{i}^{A}}\Bra{\phi_{i}^{A}}$,
where $\lambda_{i}$ is the $i$-th largest eigenvalue of $\mathbf{XX}^{\dag}$. The conservation of probability~(\ref{eq:pcv}) now implies the constraint $\sum_{i=1}^{m}\lambda_{i}=1$. The probability measure of the random coefficient matrix $\mathbf{X}$ is the Haar measure, where the entries are uniformly distributed over all the possible values satisfying the constraint~(\ref{eq:pcv}). The resulting eigenvalue density of $\mathbf{XX}^{\dag}$ is (see, e.g., Ref.~\onlinecite{Page1993}),
\begin{equation}\label{eq:fte}
f\left(\bm{\lambda}\right)=\frac{\Gamma(mn)}{c}~\delta\left(1-\sum_{i=1}^{m}\lambda_{i}\right)\prod_{1\leq i<j\leq m}\left(\lambda_{i}-\lambda_{j}\right)^{2}\prod_{i=1}^{m}\lambda_{i}^{n-m},
\end{equation}
where $\delta(\cdot)$ is the Dirac delta function and the constant
\begin{equation}\label{eq:con}
c=\prod_{i=1}^{m}\Gamma(n-i+1)\Gamma(i).
\end{equation}
The random matrix ensemble~(\ref{eq:fte}) is also known as the (unitary) fixed-trace ensemble. The considered bipartite quantum system is a fundamental model that describes the interaction between a physical object and its environment. For example~\cite{Page1993}, the subsystem $A$ is the black hole and the subsystem $B$ is the associated radiation field. In another example~\cite{Majumdar}, the subsystem $A$ is a set of spins and the subsystem $B$ represents the environment of a heat bath.

Quantum entanglement is the most fundamental feature in quantum mechanics. Quantum states that are highly entangled contain more information about different parts of the composite system. The degree of entanglement is measured by the entanglement entropy, which is a function of the eigenvalues of $\mathbf{XX}^{\dag}$. The function should monotonically increase from the separable state ($\lambda_{1}=1$, $\lambda_{2}=\dots=\lambda_{m}=0$) to the maximally-entangled state ($\lambda_{1}=\lambda_{2}=\dots\lambda_{m}=1/m$). The most well-known entanglement entropy is the von Neumann entropy
\begin{equation}\label{eq:vN}
S=-\sum_{i=1}^{m}\lambda_{i}\ln\lambda_{i},
\end{equation}
which achieves the separable state and maximally-entangled state when $S=0$ and when $S=\ln{m}$, respectively. A one-parameter generalization of the von Neumann entropy is Tsallis entropy~\cite{Tsallis98}
\begin{equation}\label{eq:T}
T=\frac{1}{q-1}\left(1-\sum_{i=1}^{m}\lambda_{i}^{q}\right),~~~~q\in\mathbb{R}\backslash\{0\},
\end{equation}
which, by l'H\^{o}pital's rule, reduces to the von Neumann entropy~(\ref{eq:vN}) when the non-zero real parameter $q$ approaches $1$. The Tsallis entropy~(\ref{eq:T}) achieves the separable state and maximally-entangled state when $T=0$ and $T=\left(m^{q-1}-1\right)/(q-1)m^{q-1}$, respectively. The Tsallis entropy enjoys certain advantage in describing the entanglement. In particular, it overcomes the inability of the von Neumann entropy to model systems with long-range interactions~\cite{Malacarne02}. The Tsallis entropy has the unique nonadditivity (also known as nonextensivity) property, whose physical relevance to quantum systems has been increasingly identified~\cite{Gell-Mann}. It also has a definite concavity for any $q$, i.e., being convex for $q<0$ and concave for $q>0$.

Statistical behavior of entanglement entropies can be understood from their moments. In particular, one is interested in higher moments beyond mean values (first moments) that govern the fluctuation of the entropies. In the literature, the first moment of von Neumann entropy $\mathbb{E}_{f}\!\left[S\right]$ (the subscript $f$ emphasizes that the expectation is taken over the fixed-trace ensemble~(\ref{eq:fte})) was conjectured by Page~\cite{Page1993}. Page's conjecture was proved independently by Foong-Kanno~\cite{Foong1994}, S\'{a}nchez-Ruiz~\cite{Ruiz1995}, Sen~\cite{Sen1996}, and Adachi-Toda-Kubotani~\cite{Adachi2009}. Recently, an expression for the variance of von Neumann entropy $\mathbb{V}\!_{f}\!\left[S\right]$ was conjectured by Vivo-Pato-Oshanin~\cite{VPO16}, which was subsequently proved by the author~\cite{Wei17}. For the Tsallis entropy, the first moment $\mathbb{E}_{f}\!\left[T\right]$ was derived by Malacarne-Mendes-Lenzi~\cite{Malacarne02}. The task of the present work is to study the variance of Tsallis entropy $\mathbb{V}\!_{f}\!\left[T\right]$.

The paper is organized as follows. In Sec.~\ref{sec:var} we derive an exact variance formula of Tsallis entropy valid for $q>-1$ in terms of finite sums of terminating hypergeometric functions. As a byproduct of the result, we provide in the Appendix another proof to the recently proved~\cite{Wei17} Vivo-Pato-Oshanin's conjecture~\cite{VPO16}. In Sec.~\ref{sec:spe} the derived variance formula is further simplified to explicit expressions in the special cases of quadratic entropy ($q=2$) and small subsystem dimensions ($m=2,3$). We summary the main results and point out possible approaches to study the higher moments in Sec.~\ref{sec:dis}.

\section{Exact Variance of Tsallis Entropy}\label{sec:var}
Similar to the case of von Neumann entropy~\cite{Page1993,Wei17}, the starting point of the calculation is to convert the moments defined over the fixed-traced ensemble~(\ref{eq:fte}) to the well-studied Laguerre ensemble, whose correlation functions are explicitly known. Before discussing the moments conversion approach, we first set up necessary definitions relevant to the Laguerre ensemble. By construction~(\ref{eq:pcv}), the random coefficient matrix $\mathbf{X}$ is naturally related to a Wishart matrix $\mathbf{YY}^{\dag}$ as
\begin{equation}\label{eq:wf}
\mathbf{XX}^{\dag}=\frac{\mathbf{YY}^{\dag}}{\tr\left(\mathbf{YY}^{\dag}\right)},
\end{equation}
where $\mathbf{Y}$ is an $m\times n$ ($m\leq n$) matrix of independently and identically distributed complex Gaussian entries (complex Ginibre matrix). The density of the eigenvalues $0<\theta_{m}<\dots<\theta_{1}<\infty$ of $\mathbf{YY}^{\dag}$ equals~\cite{Forrester}
\begin{equation}\label{eq:we}
g\left(\bm{\theta}\right)=\frac{1}{c}\prod_{1\leq i<j\leq m}\left(\theta_{i}-\theta_{j}\right)^{2}\prod_{i=1}^{m}\theta_{i}^{n-m}\e^{-\theta_{i}},
\end{equation}
where $c$ is the same as in~(\ref{eq:con}) and the above ensemble is known as the Laguerre ensemble. The trace of the Wishart matrix
\begin{equation}\label{eq:tr}
r=\tr\left(\mathbf{YY}^{\dag}\right)=\sum_{i=1}^{m}\theta_{i}
\end{equation}
follows a gamma distribution with the density~\cite{VPO16}
\begin{equation}\label{eq:r}
h_{mn}(r)=\frac{1}{\Gamma(mn)}\e^{-r}r^{mn-1},~~~~r\in[0,\infty).
\end{equation}
The relation~(\ref{eq:wf}) induces the change of variables
\begin{equation}\label{eq:cv}
\lambda_{i}=\frac{\theta_{i}}{r},~~~~i=1,\ldots,m,
\end{equation}
that leads to a well-known relation (see, e.g., Ref.~\onlinecite{Page1993}) among the densities~(\ref{eq:fte}),~(\ref{eq:we}), and~(\ref{eq:r}) as
\begin{equation}\label{eq:relation}
f\left(\bm{\lambda}\right)h_{mn}(r)\dd r\prod_{i=1}^{m}\dd\lambda_{i}=g\left(\bm{\theta}\right)\prod_{i=1}^{m}\dd\theta_{i}.
\end{equation}
This implies that $r$ is independent of each $\lambda_{i}$, $i=1,\ldots,m$, since their densities factorize.

For the von Neumann entropy~(\ref{eq:vN}), the relation~(\ref{eq:relation}) has been exploited to convert the first two moments~\cite{Page1993,Wei17} from the fixed-trace ensemble~(\ref{eq:fte}) to the Laguerre ensemble~(\ref{eq:we}). The moments conversion was an essential starting point in proving the conjectures of Page~\cite{Page1993,Ruiz1995} and Vivo-Pato-Oshanin~\cite{Wei17}. We now show that the moments conversion approach can be also applied to study the Tsallis entropy. We first define
\begin{equation}\label{eq:L}
L=\sum_{i=1}^{m}\theta_{i}^{q}
\end{equation}
as the induced\footnote{For convenience of the discussion, we define the induced entropy, which may not have physical meaning of an entropy.} Tsallis entropy of the Laguerre ensemble~(\ref{eq:we}). Using the change of variables~(\ref{eq:cv}), the $k$-th power of Tsallis entropy~(\ref{eq:T}) can be written as
\begin{equation}
T^{k}=\frac{1}{(q-1)^{k}}\left(1-\frac{L}{r^{q}}\right)^{k}=\frac{1}{(q-1)^{k}}\sum_{i=0}^{k}(-1)^{i}\binom{k}{i}\frac{L^{i}}{r^{qi}}
\end{equation}
and thus we have
\begin{equation}\label{eq:fg0}
\mathbb{E}_{f}\!\left[T^{k}\right]=\frac{1}{(q-1)^{k}}\sum_{i=0}^{k}(-1)^{i}\binom{k}{i}\mathbb{E}_{f}\!\left[\frac{L^{i}}{r^{qi}}\right].
\end{equation}
The expectation on the left hand side is computed as
\begin{eqnarray}
\mathbb{E}_{f}\!\left[\frac{L^{i}}{r^{qi}}\right]&=&\int_{\bm{\lambda}}\frac{L^{i}}{r^{qi}}f\left(\bm{\lambda}\right)\prod_{i=1}^{m}\dd\lambda_{i}\\
&=&\int_{\bm{\lambda}}\frac{L^{i}}{r^{qi}}f\left(\bm{\lambda}\right)\prod_{i=1}^{m}\dd\lambda_{i}\int_{r}h_{mn+qi}(r)\dd r\label{eq:fg1}\\
&=&\frac{\Gamma(mn)}{\Gamma(mn+qi)}\int_{\bm{\lambda}}\int_{r}L^{i}f\left(\bm{\lambda}\right)h_{mn}(r)\dd r\prod_{i=1}^{m}\dd\lambda_{i}\label{eq:fg2}\\
&=&\frac{\Gamma(mn)}{\Gamma(mn+qi)}\mathbb{E}_{g}\!\left[L^{i}\right]\label{eq:fg3},
\end{eqnarray}
where the multiplication of an appropriate constant $1=\int_{r}h_{mn+qi}(r)\dd r$ in~(\ref{eq:fg1}) along with the fact that $r^{-qi}h_{mn+qi}(r)=\Gamma(mn)h_{mn}(r)/\Gamma(mn+qi)$ lead to~(\ref{eq:fg2}) and the last equality~(\ref{eq:fg3}) is established by the change of measures~(\ref{eq:relation}). Inserting~(\ref{eq:fg3}) into~(\ref{eq:fg0}), the $k$-th moment of Tsallis entropy~(\ref{eq:T}) is written as a sum involving the first $k$ moments of the induced Tsallis entropy~(\ref{eq:L}) as
\begin{equation}\label{eq:TL}
\mathbb{E}_{f}\!\left[T^{k}\right]=\frac{\Gamma(mn)}{(q-1)^{k}}\sum_{i=0}^{k}\binom{k}{i}\frac{(-1)^{i}}{\Gamma(mn+qi)}\mathbb{E}_{g}\!\left[L^{i}\right].
\end{equation}
With the above relation~(\ref{eq:TL}), the computation of moments over the less tractable correlation functions of the fixed-trace ensemble~(\ref{eq:fte}) is now converted the one over the Laguerre ensemble~(\ref{eq:we}), which will be calculated explicitly. In particular, computing the variance $\mathbb{V}\!_{f}\!\left[T\right]=\mathbb{E}_{f}\!\left[T^2\right]-\mathbb{E}_{f}^{2}\!\left[T\right]$ requires the moments relation~(\ref{eq:TL}) for $k=1$,
\begin{equation}\label{eq:TL1}
\mathbb{E}_{f}\!\left[T\right]=\frac{1}{q-1}\left(1-\frac{\Gamma(mn)}{\Gamma(mn+q)}\mathbb{E}_{g}\!\left[L\right]\right)
\end{equation}
and $k=2$,
\begin{equation}\label{eq:TL2}
\mathbb{E}_{f}\!\left[T^{2}\right]=\frac{1}{(q-1)^{2}}\left(1-\frac{2\Gamma(mn)}{\Gamma(mn+q)}\mathbb{E}_{g}\!\left[L\right]+\frac{\Gamma(mn)}{\Gamma(mn+2q)}\mathbb{E}_{g}\!\left[L^{2}\right]\right),
\end{equation}
where the first moment relation~(\ref{eq:TL1}) has also appeared in Ref.~\onlinecite{Malacarne02}. It is seen from~(\ref{eq:TL2}) that the essential task now is to compute $\mathbb{E}_{g}\!\left[L\right]$ and $\mathbb{E}_{g}\!\left[L^{2}\right]$. Before proceeding to the calculation, we point out that in the limit $q\to1$ the derived second moments relation~(\ref{eq:TL2}) leads to a new proof to the recently proved variance formula of von Neumann entropy~\cite{Wei17} with details provided in the Appendix.

The computation of $\mathbb{E}_{g}\!\left[L\right]$ and $\mathbb{E}_{g}\!\left[L^{2}\right]$ involves one and two arbitrary eigenvalue densities, denoted respectively by $g_{1}(x_{1})$ and $g_{2}(x_{1},x_{2})$, of the Laguerre ensemble as
\begin{eqnarray}
\mathbb{E}_{g}\!\left[L\right]&=&m\int_{0}^{\infty}\!\!x_{1}^{q}~g_{1}(x_{1})\dd x_{1},\label{eq:L1}\\
\mathbb{E}_{g}\!\left[L^{2}\right]&=&m\int_{0}^{\infty}\!\!x_{1}^{2q}~g_{1}(x_{1})\dd x_{1}+m(m-1)\int_{0}^{\infty}\!\!\int_{0}^{\infty}\!\!x_{1}^{q}x_{2}^{q}~g_{2}\left(x_{1},x_{2}\right)\dd x_{1}\dd x_{2}.\label{eq:L2}
\end{eqnarray}
In general, the joint density of $N$ arbitrary eigenvalues $g_{N}(x_{1},\dots,x_{N})$ is related to the $N$-point correlation function
\begin{equation}\label{eq:cf}
X_{N}\left(x_{1},\dots,x_{N}\right)=\det\left(K\left(x_{i},x_{j}\right)\right)_{i,j=1}^{N}
\end{equation}
as~\cite{Forrester} $g_{N}(x_{1},\dots,x_{N})=X_{N}\left(x_{1},\dots,x_{N}\right)(m-N)!/m!$, where $\det(\cdot)$ is the matrix determinant and the symmetric function $K(x_{i},x_{j})$ is the correlation kernel. In particular, we have
\begin{eqnarray}
g_{1}(x_{1})&=&\frac{1}{m}K(x_{1},x_{1}),\label{eq:x}\\
g_{2}(x_{1},x_{2})&=&\frac{1}{m(m-1)}\left(K(x_{1},x_{1})K(x_{2},x_{2})-K^{2}(x_{1},x_{2})\right),\label{eq:xy}
\end{eqnarray}
and the correlation kernel $K(x_{i},x_{j})$ of the Laguerre ensemble can be explicitly written as~\cite{Forrester}
\begin{equation}\label{eq:ker}
K(x_{i},x_{j})=\sqrt{\e^{-x_{i}-x_{j}}(x_{i}x_{j})^{n-m}}\sum_{k=0}^{m-1}\frac{C_{k}(x_{i})C_{k}(x_{j})}{k!(n-m+k)!},
\end{equation}
where
\begin{equation}
C_{k}(x)=(-1)^{k}k!L_{k}^{(n-m)}(x)
\end{equation}
with
\begin{equation}
L_{k}^{(n-m)}(x)=\sum_{i=0}^{k}(-1)^{i}\binom{n-m+k}{k-i}\frac{x^i}{i!}
\end{equation}
being the (generalized) Laguerre polynomial of degree $k$. The Laguerre polynomials satisfy the orthogonality relation~\cite{Forrester}
\begin{equation}\label{eq:oc}
\int_{0}^{\infty}\!\!x^{n-m}\e^{-x}L_{k}^{(n-m)}(x)L_{l}^{(n-m)}(x)\dd{x}=\frac{(n-m+k)!}{k!}\delta_{kl},
\end{equation}
where $\delta_{kl}$ is the Kronecker delta function. It is known that the one-point correlation function admits a more convenient representation as~\cite{Ruiz1995,Forrester}
\begin{equation}\label{eq:one}
X_{1}(x)=K(x,x)=\frac{m!}{(n-1)!}x^{n-m}\e^{-x}\left(\left(L_{m-1}^{(n-m+1)}(x)\right)^{2}-L_{m-2}^{(n-m+1)}(x)L_{m}^{(n-m+1)}(x)\right).
\end{equation}
We also need the following integral identity, due to Schr{\"{o}}dinger~\cite{Schrodinger1926}, that generalizes the integral~(\ref{eq:oc}) to
\begin{eqnarray}\label{eq:Swm}
A_{s,t}^{(\alpha,\beta)}(q)&=&\int_{0}^{\infty}\!\!x^{q}\e^{-x}L_{s}^{(\alpha)}(x)L_{t}^{(\beta)}(x)\dd{x}\nonumber\\
&=&(-1)^{s+t}\sum_{k=0}^{\min(s,t)}\binom{q-\alpha}{s-k}\binom{q-\beta}{t-k}\frac{\Gamma(k+q+1)}{k!},~~~~q>-1.
\end{eqnarray}

With the above preparation, we now proceed to the calculation of $\mathbb{E}_{g}\!\left[L\right]$ and $\mathbb{E}_{g}\!\left[L^{2}\right]$. Inserting~(\ref{eq:x}) and~(\ref{eq:one}) into~(\ref{eq:L1}), and defining further
\begin{equation}
\mathcal{A}_{s,t}=A_{s,t}^{(n-m+1,n-m+1)}(n-m+q),
\end{equation}
one obtains by using~(\ref{eq:Swm}) that
\begin{eqnarray}
\mathbb{E}_{g}\!\left[L\right]&=&\frac{m!}{(n-1)!}\left(\mathcal{A}_{m-1,m-1}-\mathcal{A}_{m-2,m}\right)\\
&=&\frac{m!}{(n-1)!}\Bigg(\sum_{k=0}^{m-1}\binom{q-1}{m-k-1}^{2}\frac{\Gamma(n-m+q+k+1)}{k!}-\nonumber\\
&&\sum_{k=0}^{m-2}\binom{q-1}{m-k-2}\binom{q-1}{m-k}\frac{\Gamma(n-m+q+k+1)}{k!}\Bigg),\nonumber
\end{eqnarray}
which is valid for $q>-1$. The first moment expression in the above form has been obtained in Ref.~\onlinecite{Malacarne02} and we continue to show that it can be compactly written as a terminating hypergeometric function of unit argument. Indeed, since
\begin{equation}
\binom{q-1}{-1}\binom{q-1}{1}\frac{\Gamma(n+q)}{(m-1)!}=0,~~~~q>-1\backslash\{0\},
\end{equation}
we have
\begin{eqnarray}
\mathbb{E}_{g}\!\left[L\right]&=&\frac{m!}{(n-1)!}\sum_{k=0}^{m-1}\frac{\Gamma(n-m+q+k+1)}{k!}\left(\binom{q-1}{m-k-1}^{2}-\binom{q-1}{m-k-2}\binom{q-1}{m-k}\right)\\
&=&\frac{m!\Gamma^{2}(q)}{(n-1)!}\sum_{k=0}^{m-1}\frac{\Gamma(n+q-k)}{(m-k-1)!}\left(\frac{1}{k!^{2}\Gamma^{2}(q-k)}-\frac{1}{(k-1)!(k+1)!\Gamma(q-k+1)\Gamma(q-k-1)}\right)\nonumber\\ 
&=&\frac{m!\Gamma(q+1)\Gamma(q)}{(n-1)!}\sum_{k=0}^{m-1}\frac{\Gamma(n+q-k)}{(m-k-1)!\Gamma(q-k+1)\Gamma(q-k)k!(k+1)!}\\
&=&\frac{m!\Gamma(q+1)\Gamma(q)}{(n-1)!}\frac{\Gamma(n+q)}{(m-1)!\Gamma(q+1)\Gamma(q)}\sum_{k=0}^{m-1}\frac{(1-m)_{k}(-q)_{k}(1-q)_{k}}{(1-n-q)_{k}(2)_{k}}\frac{1}{k!}\label{eq:Lf2}\\
&=&\frac{m\Gamma(n+q)}{(n-1)!}~\!\pFq[4]{3}{2}{1-m,-q,1-q}{1-n-q,2}{1},~~~~q>-1\backslash\{0\},\label{eq:Lf3}
\end{eqnarray}
where the second equality follows from the change of variable $k\to m-1-k$, (\ref{eq:Lf2}) is obtained by repeated use of the identity
\begin{equation}\label{eq:s2h}
\Gamma(m-k)=\frac{(-1)^{k}}{(1-m)_{k}}\Gamma(m)
\end{equation}
with $(a)_{n}=\Gamma(a+n)/\Gamma(a)$ being the Pochhammer's symbol, and~(\ref{eq:Lf3}) is obtained by the series definition of hypergeometric function
\begin{equation}
\pFq[4]{p}{q}{a_{1},\dots,a_{p}}{b_{1},\dots,b_{q}}{z}=\sum_{k=0}^{\infty}\frac{(a_{1})_{k}\dots(a_{p})_{k}}{(b_{1})_{k}\dots(b_{q})_{k}}\frac{z^{k}}{k!}
\end{equation}
that reduces to a finite sum if one of the parameters $a_{i}$ is a negative integer. Inserting~(\ref{eq:Lf3}) into~(\ref{eq:TL1}), we arrive at a compact expression for the first moment of Tsallis entropy as
\begin{equation}
\mathbb{E}_{f}\!\left[T\right]=\frac{1}{q-1}\left(1-\frac{m(mn-1)!\Gamma(n+q)}{(n-1)!\Gamma(mn+q)}~\!\pFq[4]{3}{2}{1-m,-q,1-q}{1-n-q,2}{1}\right),~~~~q>-1\backslash\{0\}.
\end{equation}

We now calculate $\mathbb{E}_{g}\!\left[L^{2}\right]$. Inserting~(\ref{eq:x}) and~(\ref{eq:xy}) into~(\ref{eq:L2}), one has
\begin{equation}\label{eq:L2s}
\mathbb{E}_{g}\!\left[L^{2}\right]=I_{1}-I_{2}+\frac{m^{2}\Gamma^{2}(n+q)}{(n-1)!^{2}}~\!\pFq[4]{3}{2}{1-m,-q,1-q}{1-n-q,2}{1}^{2},
\end{equation}
where
\begin{eqnarray}
I_{1}&=&\int_{0}^{\infty}\!\!x_{1}^{2q}~K(x_{1},x_{1})\dd x_{1},\\
I_{2}&=&\int_{0}^{\infty}\!\!\int_{0}^{\infty}\!\!x_{1}^{q}x_{2}^{q}~K^{2}\left(x_{1},x_{2}\right)\dd x_{1}\dd x_{2},\label{eq:iI2}
\end{eqnarray}
and we have used the result~(\ref{eq:Lf3}) with the fact that
\begin{equation}
\int_{0}^{\infty}\!\!x^{q}K(x,x)\dd{x}=\mathbb{E}_{g}\!\left[L\right].
\end{equation}
The integral $I_{1}$ can be read off from the steps that led to~(\ref{eq:Lf3}) by replacing $q$ with $2q$ as
\begin{equation}\label{eq:I1}
I_{1}=\frac{m\Gamma(n+2q)}{(n-1)!}\pFq[4]{3}{2}{1-m,-2q,1-2q}{1-n-2q,2}{1}.
\end{equation}
Inserting~(\ref{eq:ker}) into~(\ref{eq:iI2}), and defining further (cf.~(\ref{eq:Swm}))
\begin{equation}
\mathbb{A}_{s,t}=A_{s,t}^{(n-m,n-m)}(n-m+q),
\end{equation}
the integral $I_{2}$ is written as
\begin{equation}\label{eq:I2}
I_{2}=\sum_{k=0}^{m-1}\frac{k!^{2}\mathbb{A}_{k,k}^{2}}{(n-m+k)!^2}+2\sum_{j=1}^{m-1}\sum_{i=0}^{j-1}\frac{i!j!\mathbb{A}_{i,j}^{2}}{(n-m+i)!(n-m+j)!},
\end{equation}
where by using~(\ref{eq:Swm}) and~(\ref{eq:s2h}) we obtain
\begin{eqnarray}
\mathbb{A}_{i,j}&=&\Gamma^{2}(q+1)\sum_{k=0}^{i}\frac{\Gamma(n-m+q+k+1)}{\Gamma(q-i+k+1)\Gamma(q-j+k+1)(i-k)!(j-k)!k!}\\
&=&\frac{\Gamma(n-m+q+1)}{\Gamma(q-i+1)\Gamma(q-j+1)i!j!}\sum_{k=0}^{i}\frac{(-i)_{k}(-j)_{k}(n-m+q+1)_{k}}{(q-i+1)_{k}(q-j+1)_{k}}\frac{1}{k!}\label{eq:1I2s}\\
&=&\frac{\Gamma(n-m+q+1)}{\Gamma(q-i+1)\Gamma(q-j+1)i!j!}\pFq[4]{3}{2}{-i,-j,n-m+q+1}{q-i+1,q-j+1}{1},\label{eq:1I2}
\end{eqnarray}
and similarly
\begin{equation}\label{eq:2I2}
\mathbb{A}_{k,k}=\frac{\Gamma(n-m+q+1)}{\Gamma^{2}(q-k+1)k!^{2}}\pFq[4]{3}{2}{-k,-k,n-m+q+1}{q-k+1,q-k+1}{1}.
\end{equation}
Finally, by inserting~(\ref{eq:I1}), (\ref{eq:I2}), (\ref{eq:1I2}), and (\ref{eq:2I2}) into~(\ref{eq:L2s}) we arrive at
\begin{eqnarray}\label{eq:L2F}
\mathbb{E}_{g}\!\left[L^{2}\right]&=&\frac{m^{2}\Gamma^{2}(n+q)}{(n-1)!^{2}}~\!\pFq[4]{3}{2}{1-m,-q,1-q}{1-n-q,2}{1}^{2}+\frac{m\Gamma(n+2q)}{(n-1)!}\pFq[4]{3}{2}{1-m,-2q,1-2q}{1-n-2q,2}{1}\nonumber\\
&&-\Gamma^{4}(q+1)\Gamma^{2}(n-m+q+1)\left(\sum_{i=0}^{m-1}L^{2}(i,i)+2\sum_{j=1}^{m-1}\sum_{i=0}^{j-1}L^{2}(i,j)\!\right),~q>-1\backslash\{0\},
\end{eqnarray}
where the symmetric function $L(i,j)=L(j,i)$ is
\begin{equation}
L(i,j)=\frac{\pFq[4]{3}{2}{-i,-j,n-m+q+1}{q-i+1,q-j+1}{1}}{\Gamma(q-i+1)\Gamma(q-j+1)\sqrt{i!j!(n-m+i)!(n-m+j)!}}.
\end{equation}
With the derived first two moments~(\ref{eq:Lf3}),~(\ref{eq:L2F}) and the relations~(\ref{eq:TL1}),~(\ref{eq:TL2}), an exact variance formula of Tsallis entropy is obtained.

\section{Special Cases}\label{sec:spe}
Though the derived results~(\ref{eq:Lf3}),~(\ref{eq:L2F}) may not be further simplified for an arbitrary $m$, $n$, and $q$, we will show that explicit variance expressions can be obtained in some special cases of practical relevance.

\subsection{Quadratic Entropy $q=2$}
In the special case $q=2$, the Tsallis entropy~(\ref{eq:T}) reduces to the quadratic entropy
\begin{equation}\label{eq:T2}
T=1-\sum_{i=1}^{m}\lambda_{i}^{2},
\end{equation}
which was first considered in physics by Fermi~\cite{Malacarne02}. The quadratic entropy~(\ref{eq:T2}) is the only entropy among all possible $q$ values that satisfies the information invariance and continuity criterion~\cite{Brukner09}.

By the series representations~(\ref{eq:Lf2}) and~(\ref{eq:1I2s}), the first two moments in the case $q=2$ are directly computed as
\begin{eqnarray}
\mathbb{E}_{g}\!\left[L\right]&=&mn(m+n),\\
\mathbb{E}_{g}\!\left[L^{2}\right]&=&mn\left(mn^{3}+2m^{2}n^{2}+4n^{2}+m^{3}n+10mn+4m^{2}+2\right).
\end{eqnarray}
By~(\ref{eq:TL1}) and~(\ref{eq:TL2}), we immediately have
\begin{eqnarray}
\!\!\!\!\!\!\!\mathbb{E}_{f}\!\left[T\right]&=&\frac{mn-m-n+1}{mn+1},\label{eq:ETq2}\\
\!\!\!\!\!\!\!\mathbb{E}_{f}\!\left[T^{2}\right]&=&\frac{(m-1)(n-1)}{(mn+1)(mn+2)(mn+3)}\left(m^{2}n^{2}-mn^{2}-m^{2}n+5mn-4n-4m+8\right),
\end{eqnarray}
which lead to the variance of Tsallis entropy for $q=2$ as
\begin{equation}\label{eq:q2}
\mathbb{V}\!_{f}\!\left[T\right]=\frac{2\left(m^{2}-1\right)\left(n^{2}-1\right)}{\left(mn+1\right)^2\left(mn+2\right)\left(mn+3\right)}.
\end{equation}
Finally, we note that explicit variance expressions for other positive integer values of $q$ can be similarly obtained.

\subsection{Subsystems of Dimensions $m=2$ and $m=3$}
We now consider the cases when dimensions $m$ of the smaller subsystems are small. This is a relevant scenario for subsystems consisting of, for example, only a few entangled particles~\cite{Majumdar}. For $m=2$ with any $n$ and $q$, the series representations~(\ref{eq:Lf2}) and~(\ref{eq:1I2s}) directly lead to the results
\begin{eqnarray}
\mathbb{E}_{g}\!\left[L\right]&=&\frac{q^{2}+q+2n-2}{(n-1)!}\Gamma(q+n-1),\\
\mathbb{E}_{g}\!\left[L^{2}\right]&=&\frac{2}{(n-1)!}\left(\frac{\Gamma(q+n-1)\Gamma(q+n)}{(n-2)!}+\left(2q^{2}+q+n-1\right)\Gamma(2q+n-1)\right).\label{eq:m2}
\end{eqnarray}
In the same manner, for $m=3$ with any $n$ and $q$ we obtain
\begin{eqnarray}
\mathbb{E}_{g}\!\left[L\right]&=&\frac{6n\left(q^{2}+q-3\right)+(q-2)(q-1)(q+2)(q+3)+6n^{2}}{2(n-1)!}\Gamma(q+n-2),\\
\mathbb{E}_{g}\!\left[L^{2}\right]&=&\frac{6n(q^2+q-3)+q^4+4q^3-7q^2-10q+12+6n^2}{(n-1)!(n-2)!}\Gamma(q+n-2)\Gamma(q+n-1)+\nonumber\\
&&\frac{3n(4q^2+2q-3)+8q^4+8q^3-14q^2-8q+6+3n^2}{(n-1)!}\Gamma(2q+n-2).\label{eq:m3}
\end{eqnarray}
The corresponding variances are obtained by keeping in mind the relations~(\ref{eq:TL1}) and~(\ref{eq:TL2}). For $m\geq4$, explicit variance expressions can be similarly calculated. However, it does not seem promising to find an explicit variance formula valid for any $m$, $n$, and $q$.

\section{Summary and Perspectives on Higher Moments}\label{sec:dis}
We studied the exact variance of Tsallis entropy, which is a one parameter ($q$) generalization of the von Neumann entropy. The main result is an exact variance expression~(\ref{eq:L2F}) valid for $q>-1$ as finite sums of terminating hypergeometric functions. For $q=1$, we find a short proof to the variance formula of the degenerate case of von Neumann entropy in the Appendix. For other special cases of practical importance of $q=2$, $m=2$, and $m=3$, explicit variance expressions have been obtained in~(\ref{eq:q2}),~(\ref{eq:m2}), and~(\ref{eq:m3}), respectively.

We end this paper with some perspectives on the higher moments of Tsallis entropy. In principle, the higher moments can be calculated by integrating over the correlation kernel~(\ref{eq:ker}) as demonstrated for the first two moments. In practice, the calculation becomes progressively complicated as the order of moments increases. Here, we outline an alternative path that may systematically lead to the moments of any order in a recursive manner.
We focus on the induced Tsallis entropy $L$ as defined in~(\ref{eq:L}) since the moments conversion is available~(\ref{eq:TL}). The starting point is the generating function of $L$
\begin{eqnarray}
\tau_{m}(t,q)=\mathbb{E}_{g}\!\left[\e^{tL}\right]&=&\frac{1}{c}\int_{0}^{\infty}\!\!\cdots\!\!\int_{0}^{\infty}\!\prod_{1\leq i<j\leq m}\left(\theta_{i}-\theta_{j}\right)^{2}\prod_{i=1}^{m}\theta_{i}^{n-m}\e^{-\theta_{i}+t\theta^{q}_{i}}\dd\theta_{i}\label{eq:gT}\\
&=&\frac{1}{c}\det\left(\int_{0}^{\infty}x^{i+j+n-m-2}\e^{-x+tx^{q}}\dd x\right)_{i,j=1}^{m}\label{eq:gT1},
\end{eqnarray}
which is a two-parameter ($t$ and $q$) deformation of the Laguerre ensemble~(\ref{eq:we}). Compared to the weight function $w(x)=x^{n-m}\e^{-x}$ of the Laguerre ensemble, the deformation induces a new weight function
\begin{equation}\label{eq:nw}
w(x)=x^{n-m}\e^{-x+tx^{q}},
\end{equation}
which generalizes the Toda deformation~\cite{Ismail} $w(x)=x^{n-m}\e^{-x+tx}$ with the parameter $q$. The basic idea to systematically produce the moments is to find some differential and difference equations of the generating function $\tau_{m}(t,q)$. The theory of integrable systems~\cite{Forrester} may provide the possibility to obtain differential equations for the Hankel determinant~(\ref{eq:gT1}) with respect to continuous variables $t$ and $q$ as well as difference equations with respect to the discrete variable $m$. In particular, when $q$ is a positive integer the deformation~(\ref{eq:nw}) is known as multi-time Toda deformation~\cite{Ismail}, where much of the integrable structure is known~\cite{Ismail}.


\appendix*
\section{A New Proof to the Variance Formula of von Neumann Entropy}
Vivo, Pato, and Oshanin recently conjectured that the variance of von Neumann entropy~(\ref{eq:vN}) in a random pure state~(\ref{eq:fte}) is~\cite{VPO16}
\begin{equation}\label{eq:vNv}
\mathbb{V}\!_{f}\!\left[S\right]=-\psi_{1}\left(mn+1\right)+\frac{m+n}{mn+1}\psi_{1}\left(n\right)-\frac{(m+1)(m+2n+1)}{4n^{2}(mn+1)},
\end{equation}
where
\begin{equation}\label{eq:tg}
\psi_{1}(x)=\frac{\dd^{2}\ln\Gamma(x)}{\dd x^{2}}
\end{equation}
is the trigamma function. The conjecture was firstly proved in Ref.~\onlinecite{Wei17} and here we provide another proof starting from the relation~(\ref{eq:TL2}),
\begin{equation}\label{eq:TL2a}
\mathbb{E}_{f}\!\left[T^{2}\right]=\frac{1}{(q-1)^{2}}\left(1-\frac{2\Gamma(mn)}{\Gamma(mn+q)}\mathbb{E}_{g}\!\left[L\right]+\frac{\Gamma(mn)}{\Gamma(mn+2q)}\mathbb{E}_{g}\!\left[L^{2}\right]\right).
\end{equation}
To resolve the indeterminacy in the limit $q\to 1$, we apply twice l'H\^{o}pital's rule on both sides of the above equation
\begin{equation}\label{eq:lSL}
\mathbb{E}_{f}\!\left[S^{2}\right]=\lim_{q\to 1}\mathbb{E}_{f}\!\left[T^{2}\right]=\frac{\Gamma(mn)}{2}\left(-\frac{2}{\Gamma(mn+q)}\mathbb{E}_{g}\!\left[L\right]+\frac{1}{\Gamma(mn+2q)}\mathbb{E}_{g}\!\left[L^{2}\right]\right)^{\prime\prime}\bigg|_{q=1},
\end{equation}
where $f^{\prime}=\dd f/\dd q$. Define an induced von Neumann entropy of the Laguerre ensemble~(\ref{eq:we})
\begin{equation}
R_{a}=\sum_{i=1}^{m}\theta_{i}\ln^{a}\theta_{i}
\end{equation}
with $R_{1}$ further denoted by $R=R_{1}$, the right hand side of~(\ref{eq:lSL}) can be evaluated by using the following facts
\begin{eqnarray}
\mathbb{E}_{g}\!\left[L\right]\big|_{q=1}&=&\mathbb{E}_{r}\!\left[r\right],\\
\mathbb{E}_{g}\!\left[L^{2}\right]\big|_{q=1}&=&\mathbb{E}_{r}\!\left[r^{2}\right],\\
\mathbb{E}_{g}\!^{\prime}\left[L\right]\big|_{q=1}&=&\mathbb{E}_{g}\!\left[R\right],\\
\mathbb{E}_{g}\!^{\prime\prime}\left[L\right]\big|_{q=1}&=&\mathbb{E}_{g}\!\left[R_{2}\right],\\
\mathbb{E}_{g}\!^{\prime}\left[L^{2}\right]\big|_{q=1}&=&2\mathbb{E}_{g}\!\left[rR\right],\\
\mathbb{E}_{g}\!^{\prime\prime}\left[L^{2}\right]\big|_{q=1}&=&2\mathbb{E}_{g}\!\left[R^{2}\right]+2\mathbb{E}_{g}\!\left[rR_{2}\right],
\end{eqnarray}
and the definitions of digamma function $\psi_{0}(x)=\dd\ln\Gamma(x)/\dd x$ and trigamma function~(\ref{eq:tg}) that give
\begin{eqnarray}
\Gamma^{\prime}(q)&=&\Gamma(q)\psi_{0}(q),\\
\Gamma^{\prime\prime}(q)&=&\Gamma(q)\left(\psi_{1}(q)+\psi_{0}^{2}(q)\right),
\end{eqnarray}
as
\begin{eqnarray}
\mathbb{E}_{f}\!\left[S^{2}\right]&=&\frac{1}{mn}\left(2\mathbb{E}_{g}\!\left[R\right]\psi_{0}(mn+1)+\mathbb{E}_{r}\!\left[r\right]\left(\psi_{1}(mn+1)-\psi_{0}^{2}(mn+1)\right)-\mathbb{E}_{g}\!\left[R_{2}\right]\right)+\nonumber\\
&&\frac{1}{mn(mn+1)}\Big(\mathbb{E}_{g}\!\left[R^{2}\right]+\mathbb{E}_{g}\!\left[rR_{2}\right]-4\mathbb{E}_{g}\!\left[rR\right]\psi_{0}(mn+2)-\nonumber\\
&&2\mathbb{E}_{r}\!\left[r^{2}\right]\left(\psi_{1}(mn+2)-\psi_{0}^{2}(mn+2)\right)\!\Big).\label{eq:SrR}
\end{eqnarray}
In~(\ref{eq:SrR}), the first two moments of $r$ are given by
\begin{equation}\label{eq:r12}
\mathbb{E}_{r}\!\left[r\right]=mn,~~~~\mathbb{E}_{r}\!\left[r^{2}\right]=mn(mn+1),
\end{equation}
which are obtained from the $k$-th moment expression, cf.~(\ref{eq:r}),
\begin{equation}\label{eq:rk}
\mathbb{E}_{r}\!\left[r^{k}\right]=\frac{\Gamma(mn+k)}{\Gamma(mn)}.
\end{equation}
The first two moments of the induced von Neumann entropy over the Laguerre ensemble $\mathbb{E}_{g}\!\left[R\right]$ and $\mathbb{E}_{g}\!\left[R^{2}\right]$ in~(\ref{eq:SrR}) have been computed in Ref.~\onlinecite{Ruiz1995} and Ref.~\onlinecite{Sen1996} as
\begin{equation}\label{eq:R1}
\mathbb{E}_{g}\!\left[R\right]=mn\psi_{0}(n)+\frac{1}{2}m(m+1)
\end{equation}
and in Ref.~\onlinecite{Wei17} as
\begin{eqnarray}
\mathbb{E}_{g}\!\left[R^{2}\right]&=&mn(m+n)\psi_{1}(n)+mn(mn+1)\psi_{0}^{2}(n)+m\left(m^{2}n+mn+m+2n+1\right)\psi_{0}(n)+\nonumber\\
&&\frac{1}{4}m(m+1)\left(m^2+m+2\right),\label{eq:R2}
\end{eqnarray}
respectively. The remaining task is to calculate $\mathbb{E}_{g}\!\left[rR\right]$, $\mathbb{E}_{g}\!\left[R_{2}\right]$, and $\mathbb{E}_{g}\!\left[rR_{2}\right]$ in~(\ref{eq:SrR}). This relies on the repeated use of the change of variables~(\ref{eq:cv}) and measures~(\ref{eq:relation}), which exploit the independence between $r$ and $\bm{\lambda}$. Indeed, we have
\begin{eqnarray}
\mathbb{E}_{g}\!\left[rR\right]&=&\mathbb{E}_{g}\!\left[r\sum_{i=1}^{m}r\lambda_{i}\ln\left(r\lambda_{i}\right)\right]\\
&=&\mathbb{E}_{r}\!\left[r^{2}\ln r\right]-\mathbb{E}_{g}\!\left[r^{2}S\right]\\
&=&\frac{\Gamma(mn+2)}{\Gamma(mn)}\psi_{0}(mn+2)-\mathbb{E}_{r}\!\left[r^{2}\right]\mathbb{E}_{f}\!\left[S\right]\label{eq:rR1}\\
&=&mn(mn+1)\left(\psi_{0}(n)+\frac{1}{mn+1}+\frac{m+1}{2n}\right),\label{eq:rR2}
\end{eqnarray}
where~(\ref{eq:rR1}) is obtained by~(\ref{eq:relation}) and the identity
\begin{equation}\label{eq:1eln}
\int_{0}^{\infty}\!\!\e^{-r}r^{a-1}\ln{r}\dd r=\Gamma(a)\psi_{0}(a),~~~~\operatorname{Re}(a)>0,
\end{equation}
and~(\ref{eq:rR2}) is obtained by~(\ref{eq:rk}) and the mean formula of von Neumann entropy~\cite{Page1993,Foong1994,Ruiz1995,Sen1996,Adachi2009}
\begin{equation}\label{eq:vNm}
\mathbb{E}_{f}\!\left[S\right]=\psi_{0}(mn+1)-\psi_{0}(n)-\frac{m+1}{2n}.
\end{equation}
Define a generalized von Neumann entropy to~(\ref{eq:vN}) as
\begin{equation}
S_{a}=-\sum_{i=1}^{m}\lambda_{i}\ln^{a}\lambda_{i}
\end{equation}
with $S=S_{1}$, we similarly have
\begin{eqnarray}
\mathbb{E}_{g}\!\left[rR_{2}\right]&=&\mathbb{E}_{r}\!\left[r^{2}\ln^{2}r\right]-2\mathbb{E}_{g}\!\left[r^{2}\ln rS\right]-\mathbb{E}_{g}\!\left[r^{2}S_{2}\right]\\
&=&\mathbb{E}_{r}\!\left[r^{2}\ln^{2}r\right]-2\mathbb{E}_{r}\!\left[r^{2}\ln r\right]\mathbb{E}_{f}\!\left[S\right]-\mathbb{E}_{r}\!\left[r^{2}\right]\mathbb{E}_{f}\!\left[S_{2}\right],\label{eq:rR_2}
\end{eqnarray}
where the term
\begin{equation}
\mathbb{E}_{r}\!\left[r^{2}\ln^{2}r\right]=mn(mn+1)\left(\psi_{1}(mn+2)+\psi^{2}(mn+2)\right)
\end{equation}
is obtained by the identity
\begin{equation}
\int_{0}^{\infty}\!\!\e^{-r}r^{a-1}\ln^{2}{r}\dd{r}=\Gamma(a)\left(\psi_{1}(a)+\psi_{0}^{2}(a)\right),~~~~\operatorname{Re}(a)>0,
\end{equation}
and it remains to calculate the term $\mathbb{E}_{f}\!\left[S_{2}\right]$ in~(\ref{eq:rR_2}),
\begin{eqnarray}
\!\!\!\!\!\!\!\!\!\!\!\!\!\!\!\!\mathbb{E}_{f}\!\left[S_{2}\right]&=&\int_{\bm{\lambda}}\left(-\frac{R_{2}}{r}-2S\ln r+\ln^{2}r\right)f\left(\bm{\lambda}\right)\prod_{i=1}^{m}\dd\lambda_{i}\int_{r}h_{mn+1}(r)\dd r\\
&=&-\frac{1}{mn}\mathbb{E}_{g}\!\left[R_{2}\right]-2\psi_{0}(mn+1)\mathbb{E}_{f}\!\left[S\right]+\psi_{1}(mn+1)+\psi_{0}^{2}(mn+1)\\
&=&-\frac{1}{mn}\mathbb{E}_{g}\!\left[R_{2}\right]+\psi_{1}(mn+1)-\psi_{0}^{2}(mn+1)+2\psi_{0}(mn+1)\left(\!\psi_{0}(n)+\frac{m+1}{2n}\!\right).\label{eq:S2}
\end{eqnarray}
It is seen that the term involving $\mathbb{E}_{g}\!\left[R_{2}\right]$ in the above cancels the one in~(\ref{eq:SrR}). Finally, inserting~(\ref{eq:r12}), (\ref{eq:R1}), (\ref{eq:R2}), (\ref{eq:rR2}), (\ref{eq:rR_2}), and (\ref{eq:S2}) into~(\ref{eq:SrR}), and keeping in mind the mean formula~(\ref{eq:vNm}), we prove the variance formula~(\ref{eq:vNv}) after some necessary simplification by the identities
\begin{equation}
\psi_{0}(l+n)=\psi_{0}(l)+\sum_{k=0}^{n-1}\frac{1}{l+k},~~~~\psi_{1}(l+n)=\psi_{1}(l)-\sum_{k=0}^{n-1}\frac{1}{(l+k)^2}.
\end{equation}


\providecommand{\noopsort}[1]{}\providecommand{\singleletter}[1]{#1}%

\end{document}